\documentclass[prl,twocolumn,aps,showpacs,preprintnumbers,amsmath,amssymb, superscriptaddress]{revtex4-1}
\pdfoutput=1

\usepackage{bm}
\usepackage{graphicx}
\usepackage{color}

\begin{document}

\title{Surface and Bulk Electronic Structure of the Strongly Correlated
System SmB$_{6}$ and Implications for a Topological Kondo Insulator}

\author{N. Xu}

\email{nan.xu@psi.ch}

\affiliation{Swiss Light Source, Paul Scherrer Insitut, CH-5232 Villigen PSI,
Switzerland}

\author{X. Shi}

\affiliation{Swiss Light Source, Paul Scherrer Insitut, CH-5232 Villigen PSI,
Switzerland}

\affiliation{Beijing National Laboratory for Condensed Matter Physics, and Institute
of Physics, Chinese Academy of Sciences, Beijing 100190, China}

\author{P. K. Biswas}

\affiliation{Laboratory for Muon Spin Spectroscopy, Paul Scherrer Institut, CH-5232
Villigen PSI, Switzerland}

\author{C. E. Matt}

\affiliation{Swiss Light Source, Paul Scherrer Insitut, CH-5232 Villigen PSI,
Switzerland}

\affiliation{Laboratory for Solid State Physics, ETH Z\"urich, CH-8093 Z\"urich,
Switzerland}

\author{R. S. Dhaka}

\author{Y. Huang}

\author{N. C. Plumb}

\affiliation{Swiss Light Source, Paul Scherrer Insitut, CH-5232 Villigen PSI,
Switzerland}

\author{M. Radovi\'c}

\affiliation{Swiss Light Source, Paul Scherrer Insitut, CH-5232 Villigen PSI,
Switzerland}

\affiliation{SwissFEL, Paul Scherrer Institut, CH-5232 Villigen PSI, Switzerland}

\author{J. H. Dil}

\affiliation{Physik-Institut, Universit\"at Z\"urich, Winterthurerstrauss 190, CH-8057 Z\"rich, Switzerland}

\affiliation{Swiss Light Source, Paul Scherrer Insitut, CH-5232 Villigen PSI,
Switzerland}

\author{E. Pomjakushina}

\affiliation{Laboratory for Developments and Methods, Paul Scherrer Institut,
CH-5232 Villigen PSI, Switzerland}

\author{A. Amato}

\affiliation{Laboratory for Muon Spin Spectroscopy, Paul Scherrer Institut, CH-5232
Villigen PSI, Switzerland}

\author{Z. Salman}

\affiliation{Laboratory for Muon Spin Spectroscopy, Paul Scherrer Institut, CH-5232
Villigen PSI, Switzerland}

\author{D. McK. Paul}

\affiliation{Physics Department, University of Warwick, Coventry, CV4 7AL, United Kingdom}

\author{J. Mesot}

\affiliation{Swiss Light Source, Paul Scherrer Insitut, CH-5232 Villigen PSI,
Switzerland}

\affiliation{Institut de la Matiere Complexe, EPF Lausanne, CH-1015, Lausanne,
Switzerland}

\author{H. Ding}

\affiliation{Beijing National Laboratory for Condensed Matter Physics, and Institute
of Physics, Chinese Academy of Sciences, Beijing 100190, China}

\author{M. Shi}

\email{ming.shi@psi.ch}

\affiliation{Swiss Light Source, Paul Scherrer Insitut, CH-5232 Villigen PSI,
Switzerland}

\date{\today}
\begin{abstract}
Recent theoretical calculations and experimental results suggest that
the strongly correlated material SmB$_{6}$ may be a realization of
a topological Kondo insulator. We have performed an angle-resolved
photoemission spectroscopy study on SmB$_{6}$ in order to elucidate
elements of the electronic structure relevant to the possible occurrence
of a topological Kondo insulator state. The obtained electronic structure
in the whole three-dimensional momentum space reveals one electron-like
$5d$ bulk band centred at the X point of the bulk Brillouin zone
that is hybridized with strongly correlated $f$ electrons, as well
as the opening of a Kondo bandgap ($\Delta$$_{B}$ $\sim$ 20 meV)
at low temperature. In addition, we observe electron-like bands
forming three Fermi surfaces at the center $\bar{\Gamma}$ point and boundary $\bar{X}$ point of the surface Brillouin zone. These bands
are not expected from calculations of the bulk electronic structure,
and their observed dispersion characteristics are consistent with
surface states. Our results suggest that the unusual low-temperature
transport behavior of SmB$_{6}$ is likely to be related to the pronounced
surface states sitting inside the band hybridisation gap and/or the
presence of a topological Kondo insulating state.
\end{abstract}


\maketitle
A three-dimensional (3D) topological insulator (TI) is an unusual
topological quantum state associated with unique metallic surface
states that appear within the bulk bandgap\cite{Hasan_Kane_RMP2010,SC_Zhang_RMP2011}.
Owing to the peculiar spin texture protected by time-reversal symmetry,
the Dirac fermions in TIs are forbidden from scattering due to nonmagnetic
impurities and disorder \cite{Y_Xia_NP2009,S_Souma_RPL2011}. Hence
they carry dissipationless spin current \cite{Kane_Mele_Science2006},
making it possible to explore fundamental physics, spintronics, and
quantum computing \cite{Hasan_Kane_RMP2010,SC_Zhang_RMP2011}. However,
even after extensive materials synthesis efforts \cite{H_Peng_Nat_Mater2009,J_Checkelsky_PRL2011,J_G_Analytis_Nat_Phy2010,A_Taskin_PRB2009,D_Kim_Nat_Phy2012},
impurities in the bulk of these materials make them metallic, prompting
us to search for new types of TIs with truly insulating bulks.

The 3D Kondo insulator SmB$_{6}$ may open a new route to realizing
topological surface states. SmB$_{6}$ is a typical heavy fermion
material with strong electron correlation. Localized $f$ electrons
hybridize with conduction electrons, leading to a narrow bandgap on
the order of 10 meV opening at low temperatures, with the chemical
potential lying in the gap \cite{P_Coleman_handbook2007,G_Aeppli_Z_Fisk_CCMP1992,H_Tsunetsugu_RMP1997,P_Riseborough_adv_Phy2000}.
Due to the opening of the bandgap, the conductivity changes from metallic
to insulating behavior with decreasing temperature. It saturates to
a constant value below about 1 K, which is thought to be caused by
in-gap states \cite{A_Menth_PRL1969}. Theoretical studies
have proposed that
SmB$_{6}$ may host three-dimensional topological insulating phases
\cite{M_Dzero_PRL2010,DXi_ZFang_SmB6_PRL2013}. Recently, transport experiments employing
a novel geometry \cite{J_Botimer_Transp}
showed convincing evidence of a distinct surface contribution to the
conductivity that is unmixed with the bulk contribution, suggesting
SmB$_{6}$ is an ideal topological insulator with a perfectly insulating
bulk. Point-contact spectroscopy revealed that the low-temperature
Kondo insulating state harbors conduction states on the surface, in
support of predictions of nontrivial topology in Kondo insulators
\cite{X_H_Zhang_SmB6_PRX2013}. Moreover, Lu et al. used the local
density approximation combined with the Gutzwiller method to investigate
the topological physics of SmB$_{6}$ from the first principles \cite{DXi_ZFang_SmB6_PRL2013}.
They found a nontrivial $\mathbb{Z}$$_2$ topology, indicating that SmB$_{6}$ is
a strongly correlated topological insulator. They calculated the topological
surface states, and found three Dirac cones, in contrast to most known
topological insulators. At present, topological insulators are essentially
understood within the theory of non-interacting topological theory
\cite{Hasan_Kane_RMP2010,SC_Zhang_RMP2011}. SmB$_{6}$, as one candidate
for a topological Kondo insulator, potentially offers us an opportunity
to investigate the interplay between topological states and strong
many-body interactions.

As a surface-sensitive technique, angle-resolved photoemission spectroscopy
(ARPES) is one of the best probes to investigate the surface states
and attest to their topological nature. However, previous ARPES studies
did not resolve the surface dispersion from bulk states, possibly
due to the system resolution and sample surface condition \cite{H_Miyazaki_SmB6_PRB2012,Denlinger_SmB6_PhyC2000}.
In this letter, we report high-resolution ARPES results from SmB$_{6}$
in the whole three-dimensional Brillouin zone (BZ) by tuning the incident
photon energy. Due to the high resolution of the ARPES system and
good sample quality, we are able for the first time to clearly identify
electron-like bands forming three Fermi surfaces (FS), which are distinct from the expected bulk states,
and we discuss the possible topological property of those surface states. 

High quality single crystals of SmB$_{6}$ were grown by the flux
method. ARPES measurements were performed at the Surface/Interface
Spectroscopy (SIS) beamline at the Swiss Light Source using a VG-Scienta
R4000 electron analyzer with photon energies ranging from 22 to 110
eV. The energy resolution ranged from $\sim$ 10 meV at 22 eV to $\sim$
15 meV at 110 eV. The angular resolution was around 0.2$^{o}$. Clean
surfaces for the ARPES measurements were obtained by cleaving the
crystals\emph{ in situ} in a working vacuum better than $5\times10^{-11}$
Torr. Shiny mirror-like surfaces were obtained after cleaving the
samples, confirming their high quality.

\begin{figure}[!t]
\begin{centering}
\includegraphics[width=3.4in]{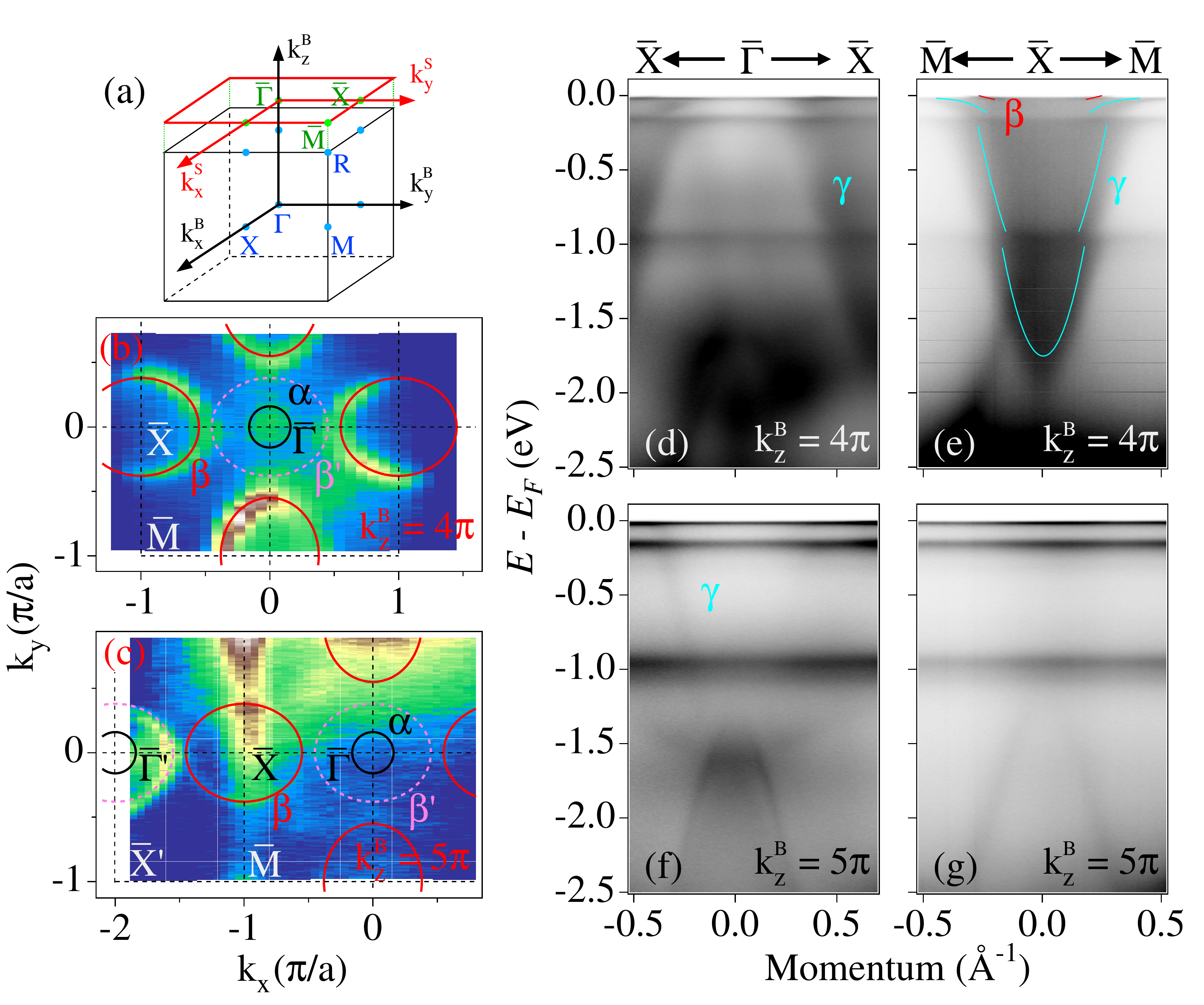} 
\par\end{centering}

\caption{\label{Fig1_FS_BS}(Color online): (a) The first Brillouin zone of
SmB$_{6}$ and the projection on the cleaving surface. High-symmetry
points are also indicated. (b),(c) Fermi surface mapping at $T$ =
17 K by integrating ARPES intensity within $E_{F}\pm5$ meV with $h\nu$
= 26 and 46 eV, corresponding $k_{z}=4\pi$ and 5$\pi$ at $\bar{\Gamma}$,
respectively. (d),(e) ARPES intensity plots at $\bar{\Gamma}$ and
$\bar{X}$ for the $k_{z}^{B}$ = 4$\pi$ plane at $T$ = 17 K. The
red and blue curves are the dispersions of the $\beta$ and $\gamma$
bands extracted by MDC fitting. (f),(g) Analogous to (d),(e), but
for the $k_{z}^{B}$ = 5$\pi$ plane. }

\label{fig1_corelevel} 
\end{figure}

\begin{figure}[!t]

\begin{centering}
\includegraphics[width=3.4in]{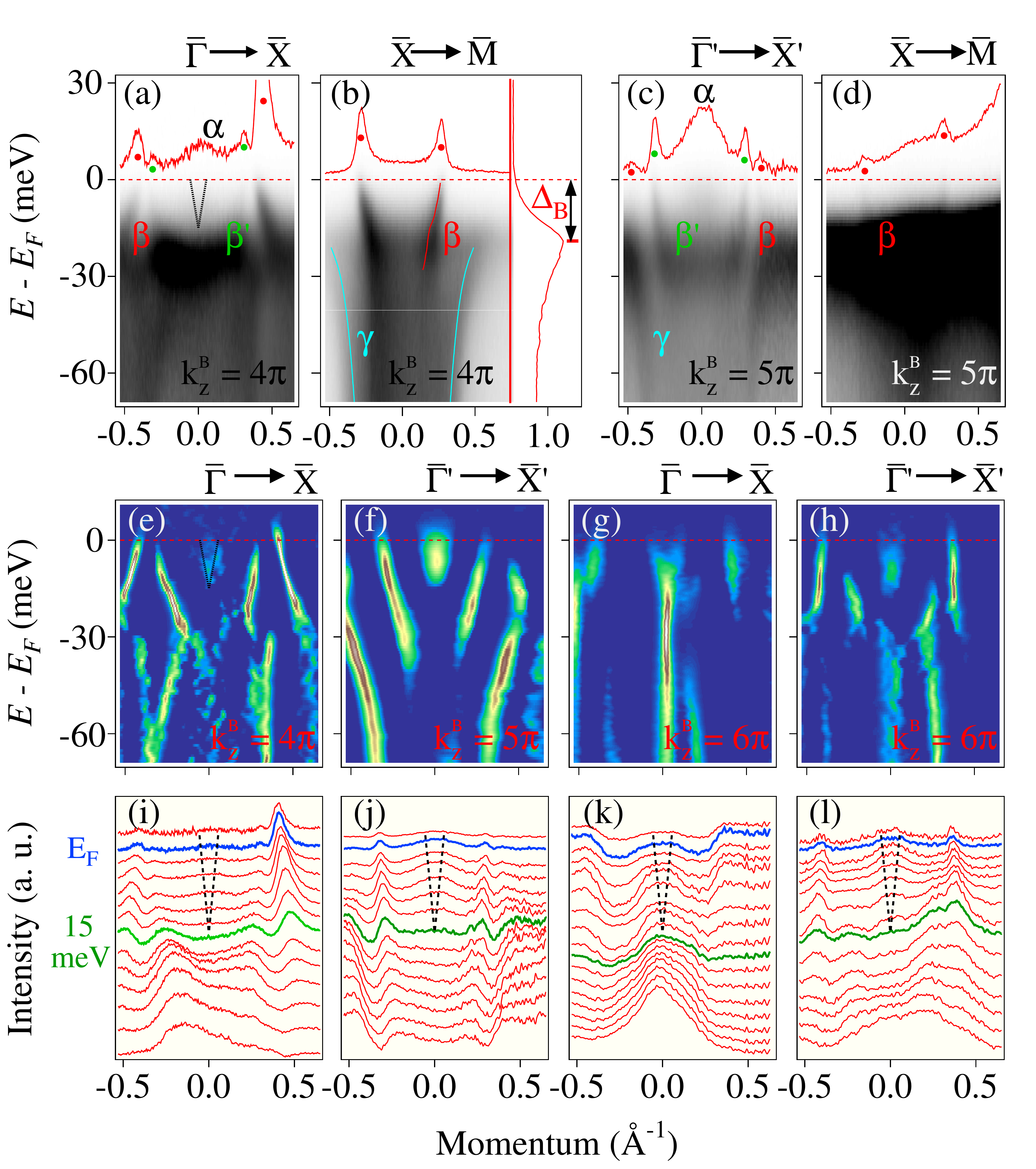} 
\par\end{centering}

\caption{\label{Fig2_FSandBS}(Color online) (a),(b) ARPES intensity plots
for cuts through $\bar{\Gamma}-\bar{X}$ and $\bar{X}-\bar{M}$, respectively,
for the $k_{z}^{B}$ = 4$\pi$ plane. Note the narrow energy window.
The data were collected at $T$ = 17 K. The red and blue curves are
dispersions for the $\beta$ and $\gamma$ bands extracted by fitting
the MDCs. The curves on the top are the MDCs taken at $E_{F}$, with
labels for the peak positions. An EDC at the location in $k$ space
marked by the red vertical line is also displayed. From this, the
Kondo bandgap is estimated to be $\Delta_{B}\sim$ 20 meV. (c),(d)
Analogous to (a),(b), but for the $k_{z}^{B}$ = 5$\pi$ plane. (e)-(g)
Plots of the curvatures of the MDC intensities along $\bar{\Gamma}-\bar{X}$
in either the first or second BZ ($ $$\bar{\Gamma}'-\bar{X}'$) evaluated
in the $k_{z}^{B}$ = 4$\pi$, 5$\pi$ and 6$\pi$ planes, respectively.
(h) MDC curvature analysis at the second BZ center $\bar{\Gamma}$'
point for the $k_{z}^{B}$ = 6$\pi$ plane. (i)-(l) Corresponding
MDC plots for (e)-(h). }

\label{fig2_FSandBS} 
\end{figure}

\begin{figure*}[!t]
\begin{centering}
\includegraphics[width=6in]{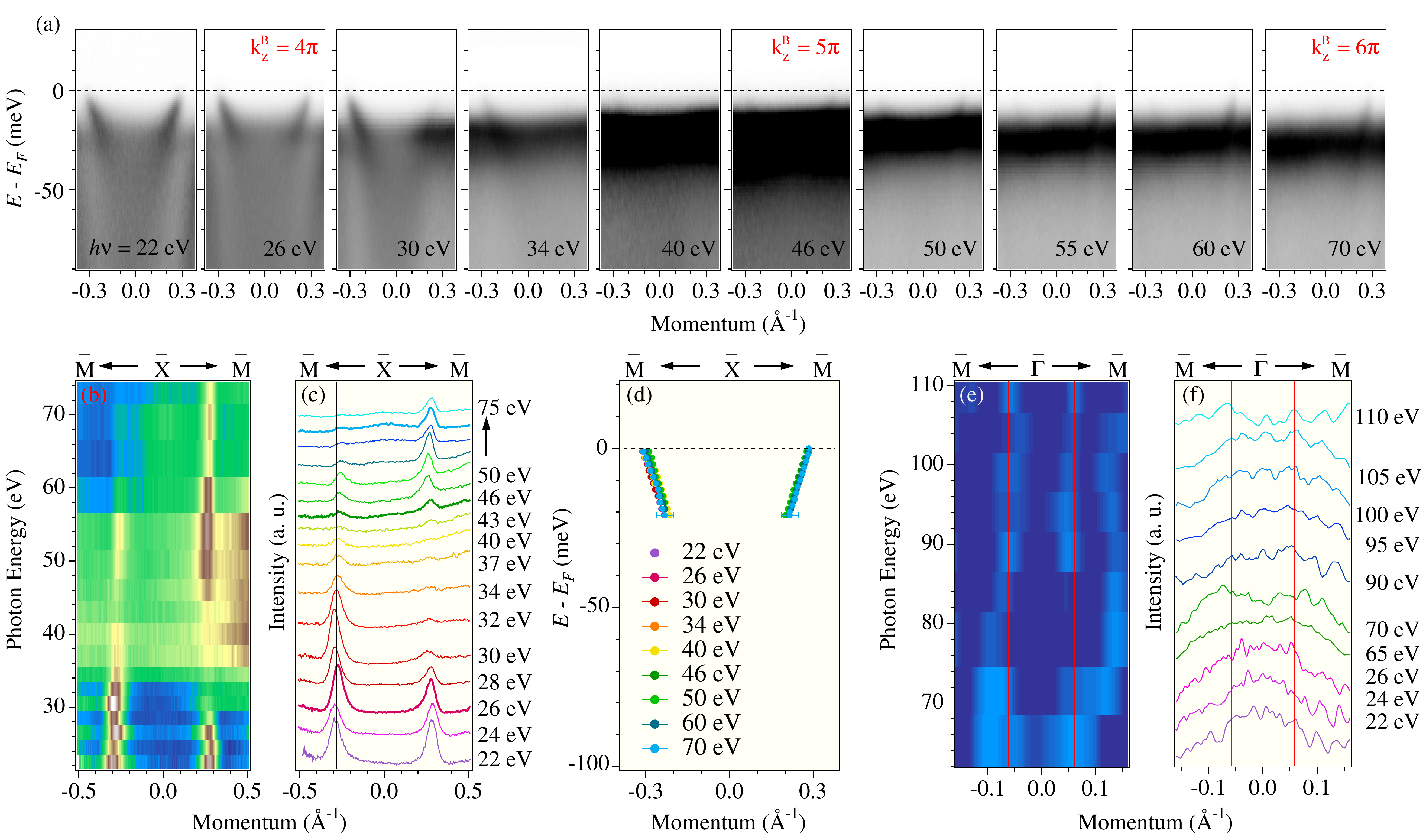} 
\par\end{centering}

\caption{\label{Fig3_kz}(Color online) (a) ARPES intensity along the $\bar{X}-\bar{M}$
direction measured at T = 17 K with various photon energies. (b) Plot
of the ARPES intensity along $\bar{X}-\bar{M}$ as a function of photon
energies from 22 eV to 75 eV, covering more than 1.5 BZs along $k_{z}^{B}$.
(c) Corresponding MDCs at $E_{F}$. (d) Extracted dispersions of the
$\beta$ band for different photon energies. (e) Plot of the curvature
of the MDC intensity along $\bar{\Gamma}-\bar{M}$ for each photon
energy. (f) Corresponding MDCs at $E_{F}$. }

\label{fig3_kz} 
\end{figure*}

Fig. \ref{Fig1_FS_BS} displays the Fermi surface and band dispersions
of SmB$_{6}$ measured at $T$ = 17 K with various photon energies,
corresponding to different $k_{z}$ points in the bulk BZ ($k_{z}^{B}$).
The first BZ of bulk SmB$_{6}$ and its projection on the cleaving
surface are shown in Fig. \ref{Fig1_FS_BS} (a), with all the high
symmetry points labeled. In Figs. \ref{Fig1_FS_BS} (b) and (c), we
plot the FS mappings obtained using $h\nu$ = 26 and 46 eV, corresponding
to approximately $k_{z}^{B}$ = 4$\pi$ and 5$\pi$, allowing direct
comparisons with previous work \cite{H_Miyazaki_SmB6_PRB2012,Denlinger_SmB6_PhyC2000}.
In making the maps, we integrated the ARPES intensity within $E_{F}\pm5$
meV. As seen in Figs. \ref{Fig1_FS_BS} (b) and (c), the same FS topology
is observed at different $k_{z}^{B}$ high symmetry points of the
bulk BZ: one small circular FS, $\alpha$, is located at the surface
BZ center $\bar{\Gamma}$ point and an additional ellipse-shaped FS,
$\beta$, is located at the surface BZ boundary $\bar{X}$ point.
We also observed a folded band $\beta$' caused by a 1$\times$2 reconstruction
of the surface, which is also observed in low-energy electron diffraction
(LEED) patterns \cite{H_Miyazaki_SmB6_PRB2012}. Figs. \ref{Fig1_FS_BS}
(d) and (e) show photoemission $E$-vs.-$k$ intensity plots at the
$\bar{\Gamma}$ and $\bar{X}$ points for the $k_{z}^{B}$ = 4$\pi$
plane at $T$ = 17 K. Similarly, data recorded for the $k_{z}^{B}=5\pi$
plane are shown in Figs. \ref{Fig1_FS_BS} (f) and (g). One can see
that the highly renormalized 4$f^{6}$ electrons form three flat bands,
located at $E_{B}$ = 960, 160 and 20 meV. One electron-like band,
$\gamma$, hybridizes with three 4$f^{6}$ bands at low temperature.
The $\gamma$ band, which is attributed to the 5$d$ orbital as suggested
by \cite{DXi_ZFang_SmB6_PRL2013}, is seen at $\bar{X}$ for $k_{z}^{B}$
= 4$\pi$ and at $\bar{\Gamma}$ for $k_{z}^{B}$ = 5$\pi$, which
in the bulk BZ are equivalent at the $X$ point,
as seen in Fig. \ref{Fig1_FS_BS} (a). This strongly three-dimensional
feature indicates that $\gamma$ is a bulk band located at the $X$
point in the bulk BZ, consistent with the theoretical calculation
\cite{DXi_ZFang_SmB6_PRL2013}.

In order to investigate the low energy excitations, band dispersions
near $E_{F}$ at $\bar{\Gamma}$ and $\bar{X}$ for the $k_{z}^{B}$
= 4$\pi$ plane are shown in Figs. \ref{Fig2_FSandBS} (a) and (b).
In addition, we plot the ARPES intensity near $E_{F}$ at the center
of the second surface BZ $\bar{\Gamma}'$ point and $\bar{X}$ point
for the $k_{z}^{B}$ = 5$\pi$ plane in Figs. \ref{Fig2_FSandBS}
(d) and (e). As seen in Figs. \ref{Fig2_FSandBS} (b) and (d), the
bulk band $\gamma$ hybridizes with the flat 4$f$ band near $E_{F}$,
leading to a Kondo bandgap. The gap size is $\Delta_{B}$ $\sim$
20 meV based on the peak position of the energy distribution curve
(EDC) taken at the position marked by the vertical red line in Fig.
\ref{Fig2_FSandBS} (b) Moreover, as seen in Figs. \ref{Fig2_FSandBS}
(b) and (d), the electron-like band $\beta$ appears inside the bandgap
and crosses $E_{F}$ at the $\bar{X}$ point, forming an ellipse-shaped
FS at the BZ boundary. For the $\beta$ band, its Fermi momentum $k_{F}$
measures 0.39 and 0.28 \AA$^{-1}$ along the $\bar{X}$-$\bar{\Gamma}$
and $\bar{X}$-$\bar{M}$ directions, respectively. The folded band $\beta$' can be seen in Figs. \ref{Fig2_FSandBS}
(a) and (c), located about at $\bar{\Gamma}$, with a folding wave
vector ($\pi$, 0) caused by the 1$\times$2 surface reconstruction.
Additionally, we observe one weak band $\alpha$ at the $\bar{\Gamma}$
point, which corresponds to the small FS at the BZ center. To better
visualize the weak $\alpha$ band, in Figs. \ref{fig2_FSandBS} (e)-(h)
we plot the curvature of the MDC intensity \cite{P_Zhang_RSI2011}
along $\bar{\Gamma}-\bar{M}$ for different photon energies approximately
corresponding to the $k_{z}^{B}$ = 4$\pi$, 5$\pi$ and 6$\pi$ planes.
The corresponding raw momentum distribution curves (MDCs) are also
plotted in Figs. \ref{fig2_FSandBS} (i)-(l). From the curvature plots,
we can see that the electron-like $\alpha$ band crosses $E_{F}$
around the $\bar{\Gamma}$ point, which can be also observed in the MDCs plots. The MDCs at $E_{F}$
in Figs. \ref{Fig2_FSandBS} (a) and (c) confirm that the $\alpha$
and $\beta$ bands indeed cross $E_{F}$.

\begin{figure}[!t]

\begin{centering}
\includegraphics[width=3.4in]{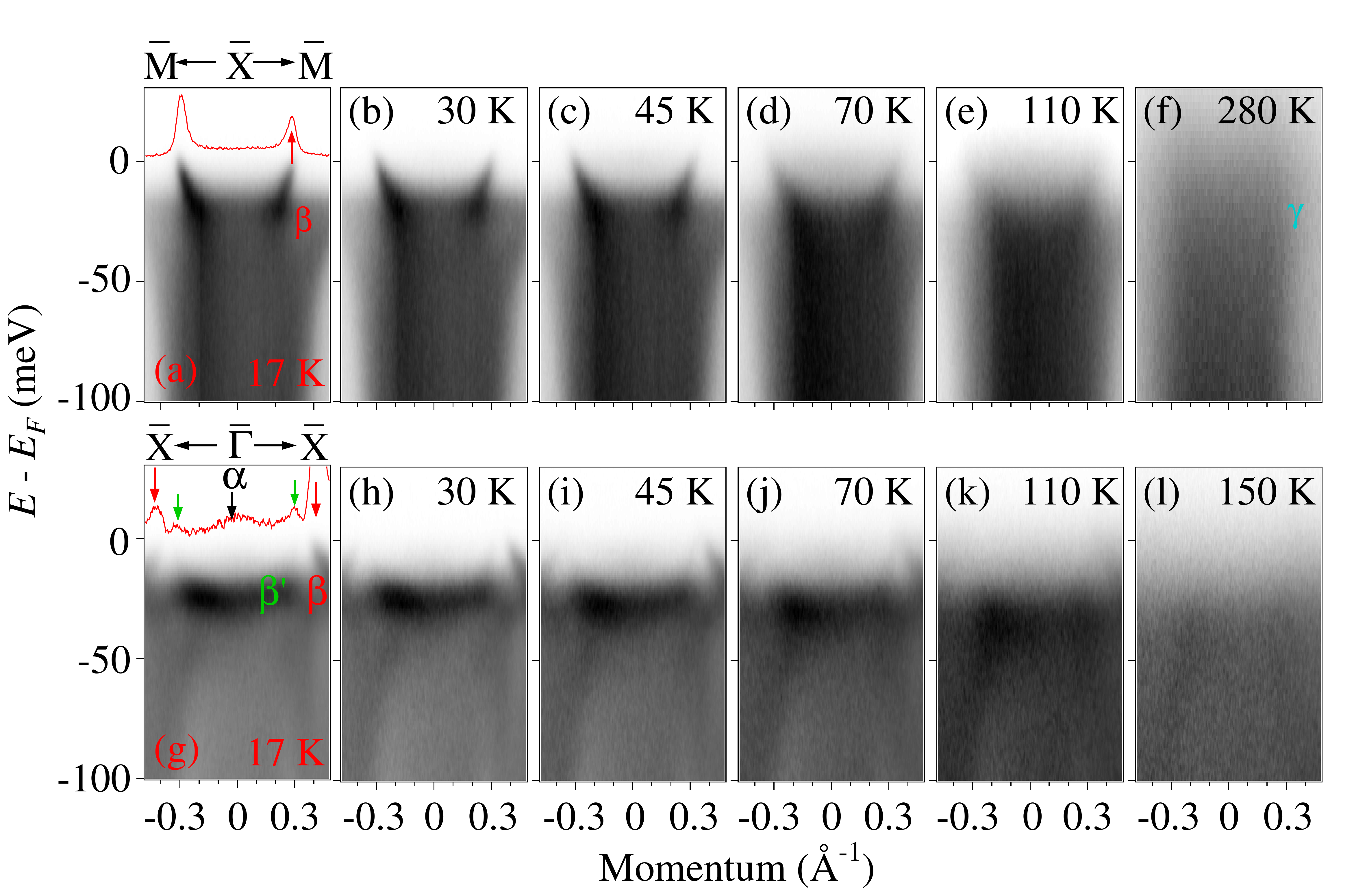} 
\par\end{centering}

\caption{\label{Fig4_T_dep}(Color online) (a)-(f) ARPES intensity plots along
$\bar{M}-\bar{X}-\bar{M}$ measured at $T$ = 17 K, 30 K, 45 K, 70
K, 110 K and 280 K, respectively. (g)-(l) ARPES intensity plots along
$\bar{X}-\bar{\Gamma}-\bar{X}$ measured at T = 17 K, 30 K, 45 K,
70 K, 110 K and 150 K, respectively.}

\label{Fig4_T_dep} 
\end{figure}

From bulk band calculations \cite{M_Dzero_PRL2010,DXi_ZFang_SmB6_PRL2013},
the in-gap bands $\alpha$ and $\beta$ are totally unexpected at
any $k_{z}^{B}$ value. However, both theoretical and experimental
results \cite{M_Dzero_PRL2010,J_Botimer_Transp,X_H_Zhang_SmB6_PRX2013,DXi_ZFang_SmB6_PRL2013}
suggest that SmB$_{6}$ exhibits metallic surface states that make
it a candidate for a strongly correlated Kondo topological insulator.
To further examine whether the in-gap states are surface or bulk bands,
we have carried out an ARPES measurement along the cut crossing the
$\bar{X}$ point for different $k_{z}^{B}$ values by tuning photon
energy. In Fig. \ref{Fig3_kz} (a) we plot ARPES spectra with $k_{||}$
oriented along the $\bar{X}$-$\bar{M}$ line taken with different
photon energies from 22 to 70 eV, which cover more than 1.5 bulk BZs
along $k_{z}^{B}$. One can see that, although the spectral weight
of the $\beta$ band varies with photon energy due to photoemission
matrix element effects, the dispersion of the $\beta$ band stays
highly fixed. MDCs at $E_{F}$ obtained with different photon energies
are plotted in Fig. \ref{Fig3_kz} (c), and the corresponding $h\nu$-$k_{||}$
FS intensity plot is shown in Fig. \ref{Fig3_kz} (b). As one immediately
recognizes from Fig. \ref{Fig3_kz} (c), the peak position of the
MDCs at $E_{F}$, which indicate the $k_{F}$ values of the $\beta$
band, are stationary with respect to $h\nu$. Thus the $\beta$ band
forms a two-dimensional FS in the $h\nu$-$k_{||}$ plane shown in Fig.
\ref{Fig3_kz} (b). In fact, when we plot the extracted dispersions
for different photon energies in Fig. \ref{Fig3_kz} (d), their linear
dispersions overlap each other within the experimental uncertainties,
demonstrating the two-dimensional nature of the $\beta$ band. This
two-dimensional feature is different from the bulk $\gamma$ band,
indicating the surface origin of the $\beta$ band. We likewise studied
the photon energy dependence of the small $\alpha$ band to check
its surface/bulk origin. While the weak intensity and shallow dispersion
make detailed quantitative analysis of the $\alpha$ band difficult,
we consistently find anomalous spectral weight at $E_{F}$ connected
to this band, independent of the photon energy. This is consistent
with a shallow 2D state that is nondispersive along $k_{z}$. In light
of the fact that the $\alpha$ band is not predicted from bulk band
structure calculations, the data strongly suggest that the $\alpha$
band, like the $\beta$ band, has a surface origin.

We have also performed temperature dependent measurements to study the evolution
of both the bulk and surface bands. Figs. \ref{Fig4_T_dep} (a)-(f)
show ARPES intensity plots at the $\bar{X}$ point measured at temperatures
ranging from 17 to 280 K. Figs. \ref{Fig4_T_dep} (g)-(i) show similar
plots at the $\bar{\Gamma}$ point measured at temperatures ranging
from 17 to 150 K. The hybridization between the 5$d$ $\gamma$ band
and 4$f$ flat band is destroyed around $T$ = 110 K. Meanwhile, the
surface bands $\alpha$ and $\beta$, as well as the folding band
$\beta'$, vanish. The temperature dependence suggests that the surface
states can only exist when the Kondo bandgap opens.

Our ARPES results demonstrate that SmB$_{6}$ is a strongly correlated
Kondo insulator with metallic surface states located inside the Kondo
bandgap. The observed surface bands at both the $\bar{\Gamma}$ and
$\bar{X}$ points show good agreement with the topologically non-trivial
surface states found in calculations \cite{DXi_ZFang_SmB6_PRL2013},
suggesting that SmB$_{6}$ is a topological Kondo insulator as predicted
theoritically \cite{M_Dzero_PRL2010,DXi_ZFang_SmB6_PRL2013}. 
Three surface bands ($\alpha$ contributes one FS at the $\bar{\Gamma}$ point and $\beta$ contributes two FSs at the $\bar{X}$ point) 
enclose an odd number of time-reversal-invariant momenta, which is a very strong indication of a topological non-trivial phase.
We also
note that no clear Dirac point is observed in our ARPES measurements;
for the $\alpha$ band, the intensity is too dim to see a potential
Dirac point clearly. One possible reason is that the cleaving surface
is the B-terminated layer, and the $\alpha$ band may originate from
Sm, making the signal very weak. For the $\beta$ band, the intensity diminishes
suddenly at $E_{B}$ $\sim$ 20 meV, corresponding to the hybridization
gap edge between $f$ and $d$ electrons, which may prohibit observing
the Dirac point formed by the bands crossing each other. Thus the
apparent absence of a clear Dirac point may be a signature of interactions
between topological surface states and the strongly correlated bulk
$f$ electrons. Such non-trivial many-body interactions have recently
been observed in other topological insulators studied by ARPES \cite{Shin_Dreesing_PRL2013}.
This hints that SmB$_{6}$ may offer an opportunity to understand
topological insulators beyond the noninteracting topological theory.

In summary, we reported high resolution ARPES results from the strongly
correlated Kondo insulator SmB$_{6}$. We first identified two anomalous
bands, $\alpha$ located at the BZ center $\bar{\Gamma}$
and two $\beta$ the BZ boundary $\bar{X}$, respectively, that are distinct from
the expected bulk band structure. While the shallow dispersion of
the $\alpha$ band prevents clear analysis of its shape as a function
of $k_{z}$, we managed to explicitly show that the $\beta$ band
is a 2D surface state. The observation of these states agrees well
with the topologically non-trivial surface states predicted by theory
calculations \cite{M_Dzero_PRL2010,DXi_ZFang_SmB6_PRL2013}. We also
observe that the $\alpha$ and $\beta$ bands disappear when the hybridization
between the bulk $\gamma$ band and the heavily correlated $f$ electrons
vanishes at high temperature. Our results uphold the possibility that
SmB$_{6}$ is a topological Kondo insulator, consistent with theoretical
calculations \cite{M_Dzero_PRL2010,DXi_ZFang_SmB6_PRL2013}.

\section*{ACKNOWLEDGMENTS}

We acknowledge Z. Fang for stimulating discussions. This work was
mainly supported by the Sino-Swiss Science and Technology Cooperation
(Project No. IZLCZ2138954). This work was also supported by grants
from MOST (2010CB923000), NSFC. This work is based in part upon research
conducted at the Swiss Light Source of the Paul Scherrer Institut
in Villigen, Switzerland, and we thank the SIS beamline staff for
their excellent support.

\bibliography{biblio_short}

\begin{thebibliography}{10}%
\makeatletter
\providecommand \@ifxundefined [1]{%
 \ifx #1\undefined \expandafter \@firstoftwo
 \else \expandafter \@secondoftwo
\fi
}%
\providecommand \@ifnum [1]{%
 \ifnum #1\expandafter \@firstoftwo
 \else \expandafter \@secondoftwo
\fi
}%
\providecommand \enquote [1]{``#1''}%
\providecommand \bibnamefont  [1]{#1}%
\providecommand \bibfnamefont [1]{#1}%
\providecommand \citenamefont [1]{#1}%
\providecommand\href[0]{\@sanitize\@href}%
\providecommand\@href[1]{\endgroup\@@startlink{#1}\endgroup\@@href}%
\providecommand\@@href[1]{#1\@@endlink}%
\providecommand \@sanitize [0]{\begingroup\catcode`\&12\catcode`\#12\relax}%
\@ifxundefined \pdfoutput {\@firstoftwo}{%
 \@ifnum{\z@=\pdfoutput}{\@firstoftwo}{\@secondoftwo}%
}{%
 \providecommand\@@startlink[1]{\leavevmode\special{html:<a href="#1">}}%
 \providecommand\@@endlink[0]{\special{html:</a>}}%
}{%
 \providecommand\@@startlink[1]{%
  \leavevmode
  \pdfstartlink
   attr{/Border[0 0 1 ]/H/I/C[0 1 1]}%
   user{/Subtype/Link/A<</Type/Action/S/URI/URI(#1)>>}%
  \relax
 }%
 \providecommand\@@endlink[0]{\pdfendlink}%
}%
\providecommand \url  [0]{\begingroup\@sanitize \@url }%
\providecommand \@url [1]{\endgroup\@href {#1}{\urlprefix}}%
\providecommand \urlprefix [0]{URL }%
\providecommand \Eprint[0]{\href }%
\@ifxundefined \urlstyle {%
  \providecommand \doi [1]{doi:\discretionary{}{}{}#1}%
}{%
  \providecommand \doi [0]{doi:\discretionary{}{}{}\begingroup
  \urlstyle{rm}\Url }%
}%
\providecommand \doibase [0]{http://dx.doi.org/}%
\providecommand \Doi[1]{\href{\doibase#1}}%
\providecommand \bibAnnote [3]{%
  \BibitemShut{#1}%
  \begin{quotation}\noindent
    \textsc{Key:}\ #2\\\textsc{Annotation:}\ #3%
  \end{quotation}%
}%
\providecommand \bibAnnoteFile [2]{%
  \IfFileExists{#2}{\bibAnnote {#1} {#2} {\input{#2}}}{}%
}%
\providecommand \typeout [0]{\immediate \write \m@ne }%
\providecommand \selectlanguage [0]{\@gobble}%
\providecommand \bibinfo [0]{\@secondoftwo}%
\providecommand \bibfield [0]{\@secondoftwo}%
\providecommand \translation [1]{[#1]}%
\providecommand \BibitemOpen[0]{}%
\providecommand \bibitemStop [0]{}%
\providecommand \bibitemNoStop [0]{.\EOS\space}%
\providecommand \EOS [0]{\spacefactor3000\relax}%
\providecommand \BibitemShut [1]{\csname bibitem#1\endcsname}%
\bibitem{Hasan_Kane_RMP2010}%
  \BibitemOpen
  \bibfield{author}{%
  \bibinfo {author} {\bibnamefont{{M. Z. Hasan and C. L. Kane}}},\ }%
  \bibfield{journal}{%
  \bibinfo {journal} {Rev. Mod. Phys.}\ }%
  \textbf{\bibinfo {volume} {82}},\ \bibinfo {pages} {3045} (\bibinfo {year}
  {2010})%
  \bibAnnoteFile{NoStop}{Hasan_Kane_RMP2010}%
\bibitem{SC_Zhang_RMP2011}%
  \BibitemOpen
  \bibfield{author}{%
  \bibinfo {author} {\bibnamefont{{X. L. Qi and S.C. Zhang}}},\ }%
  \bibfield{journal}{%
  \bibinfo {journal} {Rev. Mod. Phys.}\ }%
  \textbf{\bibinfo {volume} {83}},\ \bibinfo {pages} {1057} (\bibinfo {year}
  {2011})%
  \bibAnnoteFile{NoStop}{SC_Zhang_RMP2011}%
\bibitem{Y_Xia_NP2009}%
  \BibitemOpen
  \bibfield{author}{%
  \bibinfo {author} {\bibnamefont{{Y. Xia \emph{et al.}}}},\ }%
  \bibfield{journal}{%
  \bibinfo {journal} {Nat. Phys.}\ }%
  \textbf{\bibinfo {volume} {5}},\ \bibinfo {pages} {398} (\bibinfo {year}
  {2009})%
  \bibAnnoteFile{NoStop}{Y_Xia_NP2009}%
\bibitem{S_Souma_RPL2011}%
  \BibitemOpen
  \bibfield{author}{%
  \bibinfo {author} {\bibnamefont{{S. Souma \emph{et al.}}}},\ }%
  \bibfield{journal}{%
  \bibinfo {journal} {Phys. Rev. Lett.}\ }%
  \textbf{\bibinfo {volume} {106}},\ \bibinfo {pages} {216803} (\bibinfo {year}
  {2011})%
  \bibAnnoteFile{NoStop}{S_Souma_RPL2011}%
\bibitem{Kane_Mele_Science2006}%
  \BibitemOpen
  \bibfield{author}{%
  \bibinfo {author} {\bibnamefont{{C. L. Kane and E. J. Mele}}},\ }%
  \bibfield{journal}{%
  \bibinfo {journal} {Science}\ }%
  \textbf{\bibinfo {volume} {314}},\ \bibinfo {pages} {1692} (\bibinfo {year}
  {2006})%
  \bibAnnoteFile{NoStop}{Kane_Mele_Science2006}%
\bibitem{H_Peng_Nat_Mater2009}%
  \BibitemOpen
  \bibfield{author}{%
  \bibinfo {author} {\bibnamefont{{H. Peng \emph{et al.}}}},\ }%
  \bibfield{journal}{%
  \bibinfo {journal} {Nat. Mater.}\ }%
  \textbf{\bibinfo {volume} {9}},\ \bibinfo {pages} {225} (\bibinfo {year}
  {2009})%
  \bibAnnoteFile{NoStop}{H_Peng_Nat_Mater2009}%
\bibitem{J_Checkelsky_PRL2011}%
  \BibitemOpen
  \bibfield{author}{%
  \bibinfo {author} {\bibnamefont{{J. G. Checkelsky, Y. S. Hor, R. J. Cava and
  N. P. Ong}}},\ }%
  \bibfield{journal}{%
  \bibinfo {journal} {Phys. Rev. Lett.}\ }%
  \textbf{\bibinfo {volume} {106}},\ \bibinfo {pages} {196801} (\bibinfo {year}
  {2011})%
  \bibAnnoteFile{NoStop}{J_Checkelsky_PRL2011}%
\bibitem{J_G_Analytis_Nat_Phy2010}%
  \BibitemOpen
  \bibfield{author}{%
  \bibinfo {author} {\bibnamefont{{J. Checkelsky \emph{et al.}}}},\ }%
  \bibfield{journal}{%
  \bibinfo {journal} {Nat. Phys.}\ }%
  \textbf{\bibinfo {volume} {6}},\ \bibinfo {pages} {960} (\bibinfo {year}
  {2010})%
  \bibAnnoteFile{NoStop}{J_G_Analytis_Nat_Phy2010}%
\bibitem{A_Taskin_PRB2009}%
  \BibitemOpen
  \bibfield{author}{%
  \bibinfo {author} {\bibnamefont{{A. A. Taskin and Y. Ando}}},\ }%
  \bibfield{journal}{%
  \bibinfo {journal} {Phys. Rev. B}\ }%
  \textbf{\bibinfo {volume} {80}},\ \bibinfo {pages} {085303} (\bibinfo {year}
  {2009})%
  \bibAnnoteFile{NoStop}{A_Taskin_PRB2009}%
\bibitem{D_Kim_Nat_Phy2012}%
  \BibitemOpen
  \bibfield{author}{%
  \bibinfo {author} {\bibnamefont{{D. Kim \emph{et al.}}}},\ }%
  \bibfield{journal}{%
  \bibinfo {journal} {Nat. Phys.}\ }%
  \textbf{\bibinfo {volume} {8}},\ \bibinfo {pages} {460} (\bibinfo {year}
  {2012})%
  \bibAnnoteFile{NoStop}{D_Kim_Nat_Phy2012}%
\bibitem{P_Coleman_handbook2007}%
  \BibitemOpen
  \bibfield{author}{%
  \bibinfo {author} {\bibnamefont{{P. Coleman}}},\ }%
  \bibfield{journal}{%
  \bibinfo {journal} {Handbook of Magnetism and Advanced Magnetic Materials}\
  }%
  \textbf{\bibinfo {volume} {1}},\ \bibinfo {pages} {95} (\bibinfo {year}
  {2007})%
  \bibAnnoteFile{NoStop}{P_Coleman_handbook2007}%
\bibitem{G_Aeppli_Z_Fisk_CCMP1992}%
  \BibitemOpen
  \bibfield{author}{%
  \bibinfo {author} {\bibnamefont{{G. Aeppli and Z. Fisk}}},\ }%
  \bibfield{journal}{%
  \bibinfo {journal} {Comments Condens. Matter Phys.}\ }%
  \textbf{\bibinfo {volume} {16}},\ \bibinfo {pages} {155} (\bibinfo {year}
  {1992})%
  \bibAnnoteFile{NoStop}{G_Aeppli_Z_Fisk_CCMP1992}%
\bibitem{H_Tsunetsugu_RMP1997}%
  \BibitemOpen
  \bibfield{author}{%
  \bibinfo {author} {\bibnamefont{{H. Tsunetsugu, M. Sigrist, and K. Ueda}}},\
  }%
  \bibfield{journal}{%
  \bibinfo {journal} {Rev. Mod. Phys.}\ }%
  \textbf{\bibinfo {volume} {69}},\ \bibinfo {pages} {809} (\bibinfo {year}
  {1997})%
  \bibAnnoteFile{NoStop}{H_Tsunetsugu_RMP1997}%
\bibitem{P_Riseborough_adv_Phy2000}%
  \BibitemOpen
  \bibfield{author}{%
  \bibinfo {author} {\bibnamefont{{P. Riseborough,}}},\ }%
  \bibfield{journal}{%
  \bibinfo {journal} {Adv. Phys.}\ }%
  \textbf{\bibinfo {volume} {49}},\ \bibinfo {pages} {257} (\bibinfo {year}
  {2000})%
  \bibAnnoteFile{NoStop}{P_Riseborough_adv_Phy2000}%
\bibitem{A_Menth_PRL1969}%
  \BibitemOpen
  \bibfield{author}{%
  \bibinfo {author} {\bibnamefont{{A. Menth, E. Buehler, and T.H. Geballe}}},\
  }%
  \bibfield{journal}{%
  \bibinfo {journal} {Phys. Rev. Lett.}\ }%
  \textbf{\bibinfo {volume} {22}},\ \bibinfo {pages} {295} (\bibinfo {year}
  {1969})%
  \bibAnnoteFile{NoStop}{A_Menth_PRL1969}%
\bibitem{M_Dzero_PRL2010}%
  \BibitemOpen
  \bibfield{author}{%
  \bibinfo {author} {\bibnamefont{{M. Dzero, K. Sun, V. Galitski and P.
  Coleman}}},\ }%
  \bibfield{journal}{%
  \bibinfo {journal} {Phys. Rev. Lett.}\ }%
  \textbf{\bibinfo {volume} {104}},\ \bibinfo {pages} {106408} (\bibinfo {year}
  {2010})%
  \bibAnnoteFile{NoStop}{M_Dzero_PRL2010}%
\bibitem{DXi_ZFang_SmB6_PRL2013}%
  \BibitemOpen
  \bibfield{author}{%
  \bibinfo {author} {\bibnamefont{{F. Lu, J. Z. Zhao, H. Weng, Z. Fang and X.
  Dai}}},\ }%
  \bibfield{journal}{%
  \bibinfo {journal} {Phys. Rev. Lett.}\ }%
  \textbf{\bibinfo {volume} {110}},\ \bibinfo {pages} {096401} (\bibinfo {year}
  {2013})%
  \bibAnnoteFile{NoStop}{DXi_ZFang_SmB6_PRL2013}%
\bibitem{J_Botimer_Transp}%
  \BibitemOpen
  \bibfield{author}{%
  \bibinfo {author} {\bibnamefont{{J. Botimer \emph{et al.}}}},\ }%
  \bibfield{journal}{%
  \bibinfo {journal} {arXiv},\ \bibinfo {pages} {1211.6769}}%
   (\bibinfo {year} {2012})%
  \bibAnnoteFile{NoStop}{J_Botimer_Transp}%
\bibitem{X_H_Zhang_SmB6_PRX2013}%
  \BibitemOpen
  \bibfield{author}{%
  \bibinfo {author} {\bibnamefont{{X. H. Zhang \emph{et al.}}}},\ }%
  \bibfield{journal}{%
  \bibinfo {journal} {Phys. Rev. X}\ }%
  \textbf{\bibinfo {volume} {3}},\ \bibinfo {pages} {011011} (\bibinfo {year}
  {2013})%
  \bibAnnoteFile{NoStop}{X_H_Zhang_SmB6_PRX2013}%
\bibitem{H_Miyazaki_SmB6_PRB2012}%
  \BibitemOpen
  \bibfield{author}{%
  \bibinfo {author} {\bibnamefont{{H. Miyazaki, T. Hajiri, T. Ito, S. Kunii and
  S. I. Kimura}}},\ }%
  \bibfield{journal}{%
  \bibinfo {journal} {Phys. Rev. B}\ }%
  \textbf{\bibinfo {volume} {86}},\ \bibinfo {pages} {075105} (\bibinfo {year}
  {2012})%
  \bibAnnoteFile{NoStop}{H_Miyazaki_SmB6_PRB2012}%
\bibitem{Denlinger_SmB6_PhyC2000}%
  \BibitemOpen
  \bibfield{author}{%
  \bibinfo {author} {\bibnamefont{{J. D. Denlinger \emph{et al.}}}},\ }%
  \bibfield{journal}{%
  \bibinfo {journal} {Physica B}\ }%
  \textbf{\bibinfo {volume} {281-282}},\ \bibinfo {pages} {716} (\bibinfo
  {year} {2000})%
  \bibAnnoteFile{NoStop}{Denlinger_SmB6_PhyC2000}%
\bibitem{P_Zhang_RSI2011}%
  \BibitemOpen
  \bibfield{author}{%
  \bibinfo {author} {\bibnamefont{{P. Zhang \emph{et al.}}}},\ }%
  \bibfield{journal}{%
  \bibinfo {journal} {Rev. Sci. Instrum.}\ }%
  \textbf{\bibinfo {volume} {82}},\ \bibinfo {pages} {043712} (\bibinfo {year}
  {2011})%
  \bibAnnoteFile{NoStop}{P_Zhang_RSI2011}%
\bibitem{Shin_Dreesing_PRL2013}%
  \BibitemOpen
  \bibfield{author}{%
  \bibinfo {author} {\bibnamefont{{T. Kondo \emph{et al.}}}},\ }%
  \bibfield{journal}{%
  \bibinfo {journal} {Phys. Rev. Lett.}\ }%
  \textbf{\bibinfo {volume} {110}},\ \bibinfo {pages} {217601} (\bibinfo {year}
  {2013})%
  \bibAnnoteFile{NoStop}{Shin_Dreesing_PRL2013}%
\end{thebibliography}%

\end{document}